\documentclass[pdflatex,sn-mathphys-num]{sn-jnl}

\usepackage[numbers,sort&compress]{natbib}
\usepackage{bibunits}
\defaultbibliographystyle{sn-mathphys-num}
\defaultbibliography{sn-bibliography}
 
\makeatletter
\let\origthebibliography\thebibliography
\let\origendthebibliography\endthebibliography
\newcounter{savedbib}

\renewenvironment{thebibliography}[1]{%
  \origthebibliography{#1}%
  \setcounter{enumiv}{\value{savedbib}}%
  \setcounter{enumi}{\value{savedbib}}%
  \@ifundefined{c@NAT@ctr}{}{%
    \setcounter{NAT@ctr}{\value{savedbib}}%
  }%
}{%
  \@ifundefined{c@NAT@ctr}{%
    \setcounter{savedbib}{\value{enumiv}}%
  }{%
    \setcounter{savedbib}{\value{NAT@ctr}}%
  }%
  \origendthebibliography
}
\makeatother

\usepackage{graphicx}%
\usepackage{multirow}%
\usepackage{amsmath,amssymb,amsfonts}%
\usepackage{amsthm}%
\usepackage{mathrsfs}%
\usepackage[title]{appendix}%
\usepackage{xcolor}%
\usepackage{textcomp}%
\usepackage{manyfoot}%
\usepackage{booktabs}%
\usepackage{algorithm}%
\usepackage{algorithmicx}%
\usepackage{algpseudocode}%
\usepackage{listings}%

\theoremstyle{thmstyleone}%

\theoremstyle{thmstyletwo}%

\theoremstyle{thmstylethree}%

\usepackage{lineno}

\begin{document}

\title{Accretion Burst Crystallizes Silicates in a Planet-Forming Disk}


\author*[1,2]{\fnm{Jeong-Eun} \sur{Lee}}\email{lee.jeongeun@snu.ac.kr}

\author[1]{\fnm{Chul-Hwan} \sur{Kim}}

\author[3]{\fnm{Jaeyeong} \sur{Kim}}

\author[3]{\fnm{Seokho} \sur{Lee}}

\author[1]{\fnm{Young-Jun} \sur{Kim}}

\author[1]{\fnm{Seonjae} \sur{Lee}}


\author[1,4]{\fnm{Giseon} \sur{Baek}}

\author[5]{\fnm{Joel D.} \sur{Green}}

\author[6,7]{\fnm{Gregory J.} \sur{Herczeg}}

\author[8,9]{\fnm{Doug} \sur{Johnstone}}

\author[10]{\fnm{Klaus M.} \sur{Pontoppidan}}

\author[11]{\fnm{Yuri} \sur{Aikawa}}

\author[12]{\fnm{Yao-Lun} \sur{Yang}}

\author[13]{\fnm{Logan} \sur{Francis}}

\author[14,15]{\fnm{Mihwa} \sur{Jin}}

\author[16]{\fnm{Hyerin} \sur{Jang}}

\affil*[1]{\orgdiv{Department of Physics and Astronomy}, \orgname{Seoul National University}, \orgaddress{\street{1 Gwanak-ro}, \city{Gwanak-gu}, \postcode{08826}, \state{Seoul}, \country{Republic of Korea}}}

\affil[2]{\orgdiv{SNU Astronomy Research Center}, \orgname{Seoul National University}, \orgaddress{\street{1 Gwanak-ro}, \city{Gwanak-gu}, \postcode{08826}, \state{Seoul}, \country{Republic of Korea}}}

\affil[3]{\orgdiv{Korea Astronomy and Space Science Institute}, \orgaddress{\street{776 Daedeok-daero}, \city{Yuseong-gu}, \postcode{34055}, \state{Daejeon}, \country{Republic of Korea}}}

\affil[4]{\orgdiv{Research Institute of Basic Sciences}, \orgname{Seoul National University}, \orgaddress{\street{1 Gwanak-ro}, \city{Gwanak-gu}, \postcode{08826}, \state{Seoul}, \country{Republic of Korea}}}

\affil[5]{\orgdiv{Space Telescope Science Institute}, \orgaddress{\street{3700 San Martin Dr.}, \city{Baltimore}, \postcode{MD 21218}, \country{USA}}}

\affil[6]{\orgdiv{Kavli Institute for Astronomy and Astrophysics}, \orgname{Peking University}, \orgaddress{\street{Yiheyuan 5}, \city{Haidian Qu}, \postcode{100871}, \state{Beijing}, \country{China}}}

\affil[7]{\orgdiv{Department of Astronomy}, \orgname{Peking University}, \orgaddress{\street{Yiheyuan 5}, \city{Haidian Qu}, \postcode{100871}, \state{Beijing}, \country{China}}}

\affil[8]{\orgdiv{NRC Herzberg Astronomy and Astrophysics}, \orgaddress{\street{5071 West Saanich Road}, \city{Victoria}, \postcode{V9E 2E7}, \state{British Columbia}, \country{Canada}}}

\affil[9]{\orgdiv{Department of Physics and Astronomy}, \orgname{University of Victoria}, \orgaddress{\street{3800 Finnerty Road, Elliot Building}, \city{Victoria}, \postcode{V8P 5C2}, \state{British Columbia}, \country{Canada}}}

\affil[10]{\orgdiv{Jet Propulsion Laboratory}, \orgname{California Institute of Technology}, \orgaddress{\street{4800 Oak Grove Drive}, \city{Pasadena}, \postcode{CA 91109}, \country{USA}}}

\affil[11]{\orgdiv{Department of Astronomy, Graduate School of Science}, \orgname{The University of Tokyo}, \orgaddress{\street{7-3-1 Hongo}, \city{Bunkyo-ku}, \postcode{113-0033}, \state{Tokyo}, \country{Japan}}}

\affil[12]{\orgdiv{Star and Planet Formation Laboratory}, \orgname{RIKEN Cluster for Pioneering Research}, \orgaddress{\street{7-3-1 Hongo}, \city{Wako}, \postcode{351-0198}, \state{Saitama}, \country{Japan}}}

\affil[13]{\orgdiv{Leiden Observatory}, \orgname{Leiden University}, \orgaddress{\street{Gorlaeus Building, Einsteinweg 55}, \city{Leiden}, \postcode{2300 RA}, \state{South Holland}, \country{The Netherlands}}}

\affil[14]{\orgdiv{Astrochemistry Laboratory}, \orgname{NASA Goddard Space Flight Center}, \orgaddress{\street{Code 691}, \city{Greenbelt}, \postcode{MD 20771}, \country{USA}}}

\affil[15]{\orgdiv{Department of Physics}, \orgname{Catholic University of America}, \orgaddress{\street{WA}, \city{Washington}, \postcode{DC 20064}, \country{USA}}}

\affil[16]{\orgdiv{Department of Astrophysics/IMAPP}, \orgname{Radboud University}, \orgaddress{\street{6500 GL}, \city{Nijmegen}, \postcode{PO Box 9010}, \country{The Netherlands}}}
\begin{bibunit}
\abstract{
Crystalline silicates form at high temperatures ($>$ 900 K) \citep{Fabian2000, Hallenbeck1998}. Their presence in comets \citep{Hanner1994, Hayward2000, Wooden2002, Shinnaka2018} suggests that high-temperature dust processing occurred in the early Solar System and was subsequently transported outward to comet-forming regions. However, direct evidence for this crystallization and redistribution in Sun-like protostars has remained elusive. By comparing James Webb Space Telescope (JWST) mid-infrared spectra of the periodically bursting protostar EC 53 \citep{yhLee2020}, we detect crystalline silicate (forsterite and enstatite) emission features that appear only during the burst. The emergence of these features indicates active crystal formation via thermal annealing in the hot inner disk during the accretion burst. We also detect a nested outflow—a collimated atomic jet enclosed by slower molecular outflows, consistent with magnetohydrodynamic (MHD) wind models \citep{Pascucci2025}. This configuration provides a mechanism for outward transport of freshly crystallized silicates \citep{Giacalone2019}. Our results provide the first direct observational evidence of in-situ silicate crystallization during episodic accretion bursts in a very young star still embedded in its dense envelope. Although we do not directly detect grains transported to the outer disk, the observed trends are consistent with outward redistribution, indicating that both dust processing and transport occur during the earliest and most dynamic stages of star formation.
}

\keywords{Star Formation, Protostellar Disk, Episodic Accretion, Silicate}

\maketitle
\noindent
\textit{This is the author-accepted manuscript of an article published in
\textbf{Nature 649}, 853-858 (2026). The final version of record is available at
\href{https://doi.org/10.1038/s41586-025-09939-3}{https://doi.org/10.1038/s41586-025-09939-3}.}

\section{Main Text}\label{sec1}

The terrestrial planets and other rocky bodies in the Solar System formed from the dust material present in the Solar Nebula. 
Investigating the evolution of the mineralogy of dust grains in protoplanetary disks, the nurseries of planets, will help to unravel the processes that govern planet formation.
In the terrestrial planet-forming regions of the disk, silicates in both amorphous and crystalline forms dominate the dust content, as carbon grains are easily destroyed \citep{jeLee2010, Anderson2017}.
While amorphous silicates are widespread in the interstellar medium (ISM), 
crystalline silicates have been detected within protoplanetary disks, covering a wide range of stellar mass \citep{Bouwman2001, vanBoekel2003, vanBoekel2005, Juhasz2010, Meeus2003, Abraham2009, Olofsson2009, Furlan2011, Jang2024PDS} in star-forming regions, and also in Solar System comets \citep{Hanner1994, Hayward2000, Wooden2002, Shinnaka2018}.

Crystalline olivine (primarily Mg-rich forsterite) and crystalline pyroxenes such as enstatite represent the dominant crystalline silicate species in protoplanetary disks \citep{Bouwman2001, vanBoekel2003, vanBoekel2005, Apai2005, Olofsson2009, Juhasz2010, Jang2024PDS}, and are key building blocks of rocky bodies in planetary systems \citep{Raymond2022}. These minerals form from amorphous precursors under high-temperature conditions, typically exceeding 1000 K, reached in the inner regions of protoplanetary disks. Laboratory experiments \citep[e.g.,][]{Fabian2000, Hallenbeck1998} have demonstrated that crystallization can occur rapidly under such conditions, within minutes to hours. In contrast, crystallization is strongly suppressed at lower temperatures, below $\sim$900 K.

Surprisingly, however, spectral features of crystalline silicates have been detected in Solar System comets \citep{Hanner1994, Hayward2000, Wooden2002, Shinnaka2018} that formed in the cold, outer regions of the Solar Nebula. The Stardust mission also returned forsterite grains from comet 81P/Wild 2 \citep{Nuth2006}. Furthermore, crystalline silicates have been observed in the cold outer regions of protoplanetary disks \citep{Olofsson2009, Maaskant2014}, suggesting that additional processes (e.g., transient heating events, such as lightning \citep{Pilipp1998} and shocks \citep{Harker2002}, and/or global mixing of disk material) beyond local thermal annealing, are required to explain their presence.

Ongoing crystallization of silicates was directly observed during an outburst of the young star EX Lupi \citep{Abraham2009}, providing strong evidence that crystalline silicates can form in the surface layers of protoplanetary disks during episodic accretion events. Episodic accretion, now widely recognized as an important mode of mass assembly in low-mass star formation, is thought to occur more frequently in the early evolutionary stages, according to both theoretical and observational studies \citep{Bae2014, Park2021}. 
In addition, the inferred mean crystalline mass fractions appear to be established early during disk evolution (within less than 1 Myr) and show little variation with age \citep{Sicilia-Aguilar2009, Oliveira2011}. Understanding silicate crystallization during this early protostellar phase is, therefore, critical for constraining the thermal history of disks and the initial conditions for planet formation.

However, detecting crystalline silicates in the protostellar phase remains difficult. The thick circumstellar envelope obscures disk emission at most viewing angles, and mid-infrared (mid-IR) spectra are typically dominated by broad, amorphous silicate absorption features \citep{Quanz2007, Yang2022}. Despite the recognized importance of episodic accretion in regulating the physical and chemical evolution of protostars \citep{jeLee2019, jeLee2023, jeLee2025}, observational studies to capture a full accretion burst cycle from a single protostar in a coherent and continuous manner are rare. This is primarily due to the rarity and unpredictability of protostellar outbursts, which make time-domain studies challenging. One exception is LRLL54361, whose monthly accretion pulses have been monitored with multi-epoch mid-IR spectroscopy \citep{Muzerolle2013}. Those observations, however, did not explicitly analyze silicate mineralogy, and the source’s relatively low luminosity (0.2 to 2.7 $L_\odot$) is unlikely to drive detectable mineralogical changes during individual pulses.
 
From our monitoring observation of nearby protostars at 850 $\mu$m, we have discovered one embedded protostar, EC 53, showing cyclical bursts related to mass accretion onto the central protostar \citep{yoo2017, yhLee2020}. This periodic variability makes EC 53 a unique natural laboratory for studying how disk mineralogy evolves in response to an outburst. 

\begin{figure}[!htp]
 \centering
 \vspace{-2mm}
 \includegraphics[width=1.\textwidth]{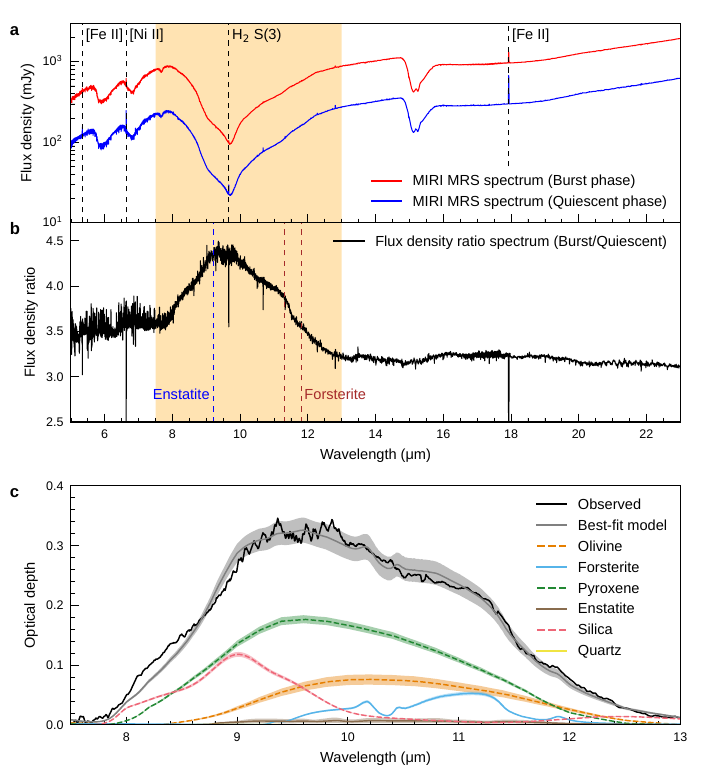}
 \vspace{-5mm}
 \caption{{\bf JWST MIRI spectra of EC 53.} {\bf (a)} The MIRI MRS spectra of EC 53 observed in the $Burst$ (red) and $Quiescent$ (blue) phases, respectively. EC 53 brightened by over a factor of 3 during the burst phase. The absorption features of various ice components and the water vapor absorption lines have been detected, with the most prominent feature being the silicate absorption at 10 $\mu$m. {\bf (b)} The flux density ratio between the burst and quiescent phases. The large enhancement around 10 $\mu$m suggests silicate emission during the burst phase. The vertical dashed lines indicate the spectral features associated with crystalline silicates. {\bf (c)} Optical depth profile (black solid line) of the silicate emission feature emerging in the 7.5 to 13 $\mu$m region (orange shading in panels a and b) during the burst phase. The combined profile (gray solid line) was composed of synthetic amorphous (olivine, pyroxene, silica) and crystalline silicates (forsterite, enstatite, and quartz). Each silicate component, combined for the best-fit model, along with its error, is presented in different colors. The 1$\sigma$ uncertainty is depicted with the shaded region.}
\label{fig:Final_Fig1}
\end{figure}

EC 53 (V371 Ser) is an embedded Class I protostar located in the Serpens Main star-forming region at a distance of 436 pc \citep{Ortiz-Leon2017}.
Cyclic variations of $\sim$1.5 years have been discovered across near-IR to sub-mm wavelengths \citep{Hodapp1999,Hodapp2012,yhLee2020,Francis2022} (Extended Data Fig.~\ref{fig:Final_FigExt1}). 
During the rise time of $\sim$0.1 year \citep{yhLee2020}, the internal luminosity increases by a factor
of $\sim$3.3 (from 6 to 20 $L_\odot$) \citep{Baek2020}, heating the disk at all radii. 
The minimum stellar mass is $\sim$0.3$\pm$0.1\, $M_\odot$, and the size and inclination of the disk are $\sim$100 au and 34.8$\pm2.1^{\circ}$\citep{shlee2020}, respectively.
Accretion variability drives the periodic brightening and fading \citep{Baek2020}, possibly triggered by interactions between the disk and a nearby ($<$ a few au) companion, potentially a protoplanet \citep{Bonnell1992,Nayakshin2012}. 
EC 53 also has a well-developed outflow from northwest (redshifted) to southeast (blueshifted); the southeast part of the disk faces us through the blueshifted outflow cavity, along which the scattered emission has been observed in K band \citep{Hodapp2012} (see Extended Data Fig.~\ref{fig:Final_FigExt2}).

Due to its periodicity, JWST observations of EC 53 could be scheduled to capture both the quiescent and burst phases. As expected, the JWST Mid-Infrared Instrument (MIRI) spectra of EC 53 exhibit strong silicate absorption features around 10 $\mu$m and 18 $\mu$m during both quiescent and burst phases, as shown in Fig.~\ref{fig:Final_Fig1}a (also see Methods). However, the flux density ratio between the two epochs is significantly high at the silicate bands of 10 $\mu$m compared to the overall continuum, as shown in Fig.~\ref{fig:Final_Fig1}b, indicating the emergence of new silicate emission features during the burst. The 10 $\mu$m silicate emission feature arises under physical conditions characteristic of the hot inner disk surface \citep{Jang2024Model}.

The optical depth profile was derived from the newly emerging silicate emission features around 10 $\mu$m (see Methods for detailed procedures) and fitted with theoretical opacities of various silicate species (Fig.~\ref{fig:Final_Fig1}c). 
Reproducing the observed silicate emission features in EC 53 requires both crystalline and amorphous components. We synthesized the silicate profiles of the amorphous olivine (MgFeSiO$_4$), pyroxene (Mg$_{0.7}$Fe$_{0.3}$SiO$_3$), and silica (SiO$_{2}$), along with crystalline forsterite (Mg$_{1.9}$Fe$_{0.1}$SiO$_4$), enstatite (Mg$_{0.96}$Fe$_{0.04}$SiO$_3$), and quartz (SiO$_{2}$), employing the $optool$ \citep{Dominik2021} and assuming 
a size distribution from 0.1 to 2 $\mu$m following a power-law index of 3.5. A bigger grain size distribution of 1 to 5 $\mu$m was required for enstatite to reproduce the observed emission adequately.
As expected from their relative activation energies for crystallization, the crystalline to amorphous mass fraction is the highest for forsterite (33\%), followed by enstatite (7.4\%) and quartz ($<$ 0.5\%) (see Extended Data Table~\ref{<tab:Final_tab>} in Methods). EC 53 shows higher crystallinity than typical T Tauri stars (average 16\%; \citep{Olofsson2010}).

The observed crystals must be hot ($>$900 K) and formed in situ during the outburst, since the 18 $\mu$m emission feature is well reproduced only by amorphous olivine, with negligible contribution from crystalline species (see Extended Data Fig.~\ref{fig:Final_FigExt6}). The 18 $\mu$m band generally traces cooler regions of the disk surface than the 10 $\mu$m band, because it corresponds to longer-wavelength vibrational modes of silicates that dominate at lower dust temperatures \citep[e.g.,][]{Abraham2009, Lu2022}. If the observed crystalline emission at 10 $\mu$m had instead originated from pre-existing silicates, then crystalline features should also have appeared at 18 $\mu$m, since the crystallinity remains similar even in the colder outer disk \citep{Olofsson2010}.

\begin{figure}[!htp]
 \centering
 \vspace{-2mm}
 \includegraphics[width=1.\textwidth]{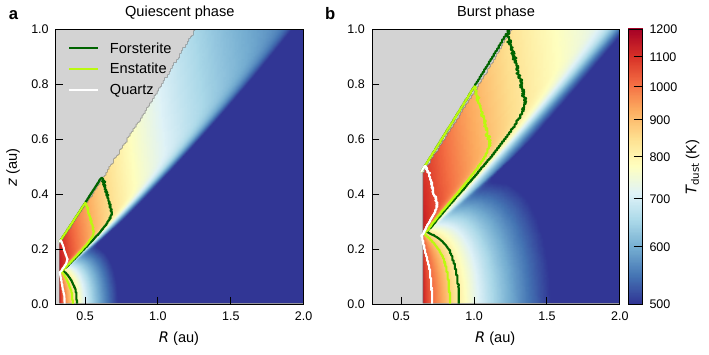}
 \vspace{-5mm}
 \caption{\textbf{Temperature distribution and crystallization regions of EC 53.} The crystallization regions of quartz (white line), enstatite (greenish-yellow line), and forsterite (dark green line) in the EC 53 disk during the quiescent (left) and burst (right) phases. These regions were calculated based on the disk model used to fit the spectral energy distribution of EC 53 \citep{Baek2020}. The inner boundary of the disk is set at the radius corresponding to the dust sublimation temperature (1200 K) and, therefore, extends to a significantly larger radius during the burst phase than in the quiescent phase. The color images in both panels represent the distribution of dust temperatures. The size of the crystallization region varies significantly depending on the activation energy for crystallization. 
The residence timescale was calculated using a viscosity parameter $\alpha$ of 0.01 for both the quiescent and burst phases in this figure. If an $\alpha$ value of 0.001 is adopted for the quiescent phase, the forsterite crystallization region at the disk surface extends slightly, from 0.6 au to 0.7 au.}
\label{fig:Final_Fig2}
\end{figure}

Fig.~\ref{fig:Final_Fig2} illustrates the crystallization regions of three crystalline silicate types overlaid on the dust temperature distribution within the inner hot disk of EC 53 \citep{Baek2020} (see Methods). These crystallization regions are identified by comparing the crystallization timescale with the residence timescale of dust grains \citep{Jang2024Model}. The residence timescale is defined as the duration a dust grain remains at a specific location in the disk, and it depends on the central stellar mass, the disk scale height, and the viscosity parameter \citep{Jang2024Model}.
Dust grains crystallize if they remain at a specific local temperature for a duration exceeding their crystallization timescale, i.e., if the crystallization timescale is shorter than the residence timescale at the location. 

The crystallization timescale depends on the dust temperature and the activation energy of grains with a particular composition \citep{Jang2024Model}.
Among the silicates, forsterite exhibits the largest crystallization region, as its activation energy \citep{Fabian2000} is the lowest. Consequently, its crystallization region extends into lower-temperature regions of the disk, as indicated by the dark green contour. Crystallization regions are more extended on the disk surface than in the midplane because high-energy photons are more effectively blocked by the high dust density in the midplane. 
At the outer edge of the crystallization region (the dark green contour in Fig.~\ref{fig:Final_Fig2}), the residence timescale is shorter than 100 days, the approximate duration of a burst phase.

During the burst phase, the temperature exceeds the grain evaporation threshold ($>$1200 K; \citep{Lenzuni1995_1200K}) in those parts of the disk where crystallization can occur in the quiescent phase ($<$0.65 au; Fig.~\ref{fig:Final_Fig2}a). As a result, most crystalline silicates formed during quiescence are evaporated during the burst. In the disk midplane, where planets and comets form, silicates cannot crystallize beyond 0.9 au, even in the burst phase. However, crystalline silicates produced on the disk surface can be transported to the midplane.
The vertical mixing timescale of dust grains is shorter than ten years at $r < 1.3$ au if a viscosity parameter exceeds $\alpha > 0.01$ (see Methods and Extended Data Fig.~\ref{fig:Final_FigExt8}). 
Thus, the fraction of crystalline silicates in the disk midplane is expected to increase over time. By contrast, radial drift to larger radii will be negligible, as its timescale is much longer than the accretion variability period in EC 53 \citep{Jang2024Model}.

\begin{figure}[!htp]
 \centering
 \vspace{-2mm}
 \includegraphics[width=1.\textwidth]{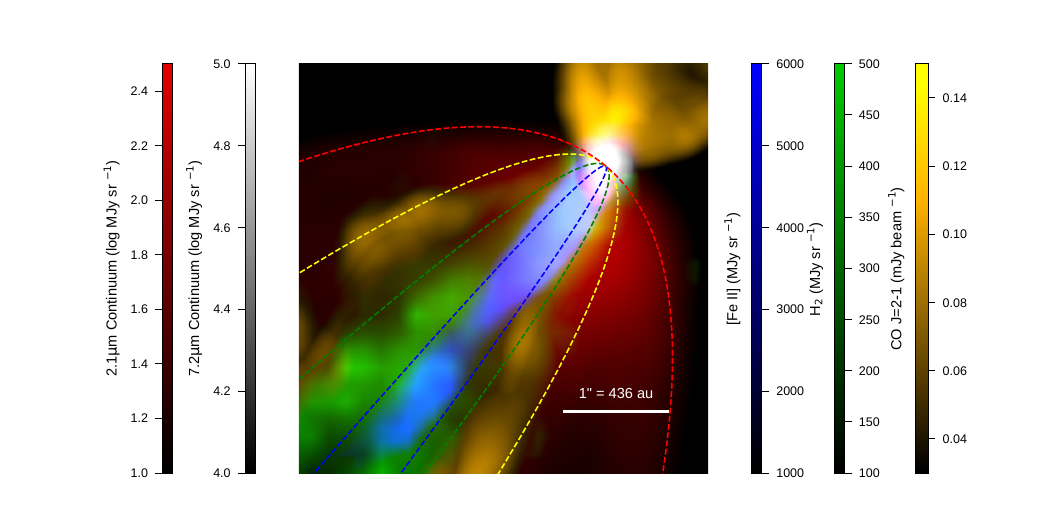}
 \vspace{-5mm}
 \caption{\textbf{The nested jet/outflow structure of EC 53.} Each color highlights a different component of the outflow. Red traces the cavity seen in scattered light at 2.1 $\mu$m, as detected by NIRCam \citep{Green2024}. Yellow indicates the cold CO J=2-1 outflow observed by ALMA (project code: 2022.1.00800.S). Green shows the H$_2$ emission at 6.91 $\mu$m (S(5) line, distinct from the line marked in Fig.~\ref{fig:Final_Fig1}a), and blue marks the [Fe II] emission at 5.34 $\mu$m. White represents the continuum at 7 $\mu$m. The line emission shows the peak intensity after continuum subtraction (Lee et al., submitted). The envelope fully obscures the redshifted outflow in the northeast at mid-IR wavelengths, whereas the submillimeter redshifted CO outflow is clearly detected. The terminal velocities of the [Fe II], H$_2$, and CO J=2-1 lines are $\sim$100, 50, and 10 km s$^{-1}$, respectively, from the inner to the outer regions. The spatial scale is indicated in the lower-right corner, and scale bars for the different emission components are also shown.}
\label{fig:Final_Fig3}
\end{figure}

\begin{figure}[!htp]
 \centering
 \vspace{-2mm}
 \includegraphics[width=1.\textwidth]{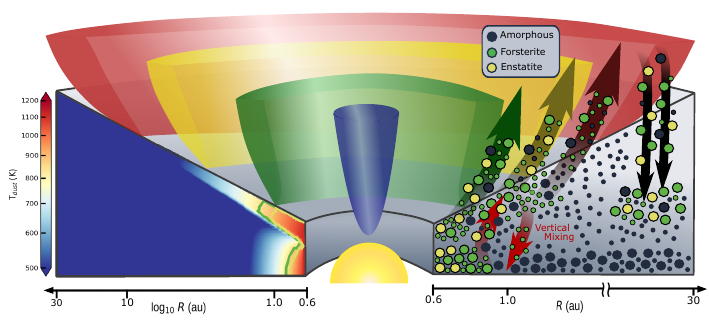}
 \vspace{-5mm}
 \caption{\textbf{Silicate crystallization and redistribution via MHD disk winds.} The left side displays the two-dimensional temperature distribution of the disk during the burst phase, highlighting the crystallization regions for each silicate species: greenish-yellow for enstatite and green for forsterite (see Fig.~\ref{fig:Final_Fig2}). The right side of the disk illustrates the crystallization and subsequent mixing of silicates. Spheres colored green and greenish-yellow represent forsterite and enstatite, respectively. These crystallized silicates can be uplifted and transported to the outer regions of the disk, the comet-forming zone, by MHD disk winds, which can drive the nested morphology of atomic jet and molecular outflows (see Fig.~\ref{fig:Final_Fig3}), presented by vertical layers. Efficient vertical mixing also increases the fraction of crystalline silicates in the disk midplane with a mixing timescale of a few years at $\sim$1 au. Note that a logarithmic scale is used on the left side of the disk, and a linear scale on the right.}
\label{fig:Final_Fig4}
\end{figure}

However, magnetically driven winds can lift crystalline silicates, annealed in the inner regions of protoplanetary disks, and transport them outward to colder disk regions. 
The X-wind model, which could launch dust from the edge of the stellar magnetosphere, was originally proposed to explain calcium-aluminum-rich inclusions (CAIs) surrounded by thick chondrule mantles in chondritic meteorites \citep{Shu1996, Shu2001}. However, in this model, the temperature at the launching radius may exceed the grain sublimation threshold. In contrast, the MHD wind model offers a more plausible mechanism for uplifting crystalline silicates over extended radii beyond the dust sublimation radius \citep{Giacalone2019} (see Methods).
Observations of EX Lup provide direct evidence for such a process: within six months after its outburst, crystalline features at longer wavelengths strengthened while the 10 $\mu$m crystallinity weakened, consistent with rapid outward transport of crystals from the inner to outer disk, most likely driven by a disk wind \citep{Juhasz2012}.
This MHD disk wind is characterized by a morphology consisting of a collimated central atomic jet nested within a wider, lower-velocity H$_2$ outflow that has been detected toward several disk sources \citep{Pascucci2025}, thanks to the unprecedented sensitivity and resolution of JWST.
Our JWST observations of EC 53 also exhibit this distinctive nested morphology of atomic jet and molecular outflow structures (Fig.~\ref{fig:Final_Fig3}), providing strong evidence that crystalline silicates formed during episodic accretion bursts can be transported to comet-forming regions beyond the hot inner zone where they are created, as illustrated in Fig.~\ref{fig:Final_Fig4}.

Episodic accretion can thermally process silicates in both embedded protostars and more evolved, disk-only systems. A useful comparison is EX Lup, whose rare but large outbursts revealed silicate processing in disk surface emission. By contrast, EC 53 is embedded, with a mid-IR spectrum dominated by circumstellar envelope absorption. Yet, even in this embedded phase, we still detect burst–quiescent cycle–linked changes in the silicate profile, implying that burst-driven heating and subsequent vertical mixing operate even during the embedded phase. The bursts of EC 53 are moderate and recurrent, possibly companion-modulated or triggered at the inner edge of the dead zone \citep[e.g.,][]{Cecil2024}, whereas EX Lup exhibits larger, less frequent events. A new JWST study of the older, Class II binary DQ Tau \citep{kospal2025} provides an instructive contrast: although this system also undergoes periodic periastron-driven brightening, the enhanced luminosities are too small to produce new crystalline silicates, and the spectral shape remains constant over time.
Because a substantial fraction of Sun-like stars are in multiple systems and close/eccentric binaries can enhance periastron accretion \citep{Zagaria2022, kospal2025}, binarity may modulate burst cadence and geometry. However, the mineralogical outcome should primarily depend on the time-integrated heating of the inner disk rather than on the specific trigger. Taken together, EC 53 and EX Lup thus outline a continuum of burst modes across evolutionary stages, indicating that both repeated moderate bursts and occasional large events can drive inner-disk silicate processing.

In contrast, crystalline silicate features have not been detected in the disk surface of FU Orionis objects \citep{Green2006, Quanz2007, Kospal2020}, which are young stellar objects undergoing prolonged, high-luminosity accretion outbursts, likely due to either the high optical depth of the emitting region or the amorphization of grains by energetic ions \citep{Glauser2009}. 
In addition to this non-detection of crystalline silicates in other eruptive young stars, the relatively low frequency of such large-scale outbursts during protostellar evolution \citep{Fischer2023} further suggests that the dominant pathway for silicate crystallization may instead be through shorter, more moderate bursts, which are more prevalent.
The JCMT Transient Survey over 6 years finds that more than one-third of the protostars are variable \citep{Mairs2024}, and the NEOWISE light curves show that more than half of protostars detected at mid-IR wavelengths are variable \citep{Park2021}.
Our discovery of crystallization occurring during a burst phase of an embedded protostar, EC 53, therefore, implies that the proto-Sun likely could have experienced a similar sequence of episodic accretion events early in its evolution. These bursts would have produced crystalline silicates in the hot, sub-au inner disk and transported them outward to the cold, comet-forming regions at tens of au via MHD disk winds.

\renewcommand\refname{References}
\putbib
\end{bibunit}

\makeatletter
\DeclareRobustCommand{\silentcitenum}[1]{%
  \begingroup
    \let\citation\@gobble
    \@ifundefined{NAT@wrout}{}{%
      \let\NAT@wrout\@gobble
    }%
    \citenum{#1}%
  \endgroup
}
\makeatother

\clearpage

\begin{bibunit}
\section{Methods}\label{methods}

\subsection{Time constrained JWST observations}\label{obs}
To investigate changes in infrared spectral features in response to the periodic brightness variation of EC 53, time-constrained NIRSpec and MIRI observations were conducted during both the quiescent and burst phases (PID: 3477, PI: Jeong-Eun Lee). The K-band monitoring observations [\silentcitenum{yhLee2020}] were used to predict the quiescent and burst phases during the JWST Cycle 2 period.
Because the source shows a long-term gradual brightness variation in addition to the 530-day periodic variability, we optimized the offsets to align the phase curves [\silentcitenum{yhLee2020}]. We then produced a phase-folded diagram with a 530-day period (see Figure 2 of [\silentcitenum{yhLee2020}]) to produce Extended Data Fig.~\ref{fig:Final_FigExt1}. The JWST observations for the quiescent and burst phases were carried out on October 5, 2023, and May 10, 2024, respectively, as indicated by the blue and red vertical dashed lines in Extended Data Fig.~\ref{fig:Final_FigExt1}. In this paper, we present only the MIRI spectra.

For the MIRI Medium Resolution Spectroscopy (MRS) mode observation, science data were obtained in the 4.9–27.9 $\mu$m range (R $\sim$ 3700–1300) to capture ice absorption features from abundant ice components, the bulk of complex organic molecules, as well as thermal absorption and emission from the disk and hot inner envelope. The MRS consists of four channels, each further divided into three sub-bands: Short (A), Medium (B), and Long (C).

For the quiescent phase observation, the FASTR1 readout pattern was used, with 35, 45, and 40 groups for the A, B, and C sub-bands, respectively, and 2 integrations per band. For the burst phase observation, the FASTR1 readout pattern was used, with 18, 16, and 16 groups for the A, B, and C sub-bands, respectively, and 5 integrations per band. A 4-point dither pattern was applied for both observations. The total exposure times were 2697 and 2908 seconds for the quiescent and burst phases, respectively.
2MASS-J18295360+0117017 was used for the target acquisition (TA) observation in the FND filter with the FAST readout pattern. A dedicated background observation was conducted to ensure proper calibration of the MIRI data for EC 53.

Extended Data Fig.~\ref{fig:Final_FigExt2} presents a larger field-of-view NIRCam image of EC 53, comparing it with the coverage of our MIRI observations (middle and right panels). The MIRI emission peak aligns well with the ALMA submillimeter continuum position, and a clear brightening is detected during the burst phase.

\subsection{Data Reduction}\label{data}

The observations from both phases were reduced using the JWST Science Calibration Pipeline version 1.18.0 \citep{Bushouse2023} with the Calibration Reference Data System \citep{Greenfield2016} context jwst\_1364.pmap. The raw data were first processed through the Detector1Pipeline 
to produce slope images with detector-level corrections under the default parameters, along 
with minor additional corrections for electromagnetic interference noise patterns and reference pixels.
Following this, the Spec2Pipeline was applied for the instrumental correction, including flat-fielding and flux calibration. The science and dedicated background observations were reduced individually. Then, master background subtraction was conducted in the Spec3Pipeline, 
producing the final 3D science data cubes for each channel and sub-band.

\subsection{Astrometry $\&$ Spectral Extraction}\label{astro}
To directly analyze the changes between the quiescent and burst phases in EC 53, the spectra from both phases must be extracted from the exact same region. However, a spatial offset exists between the quiescent and burst phase observations across all MIRI datasets.
The NIRCam F210M image (PID: 1611; PI: Pontoppidan, Klaus M.) and the ALMA submillimeter continuum image (project code: 2022.1.00800.S; PI: Lee, Seokho) show that the bright region in the NIRCam F210M image aligns well with the ALMA continuum position.
Therefore, we performed astrometric corrections for both phase observations using the ALMA continuum position as a reference. First, we created integrated intensity maps for each MIRI Channel (CH1, CH2, CH3, and CH4) in both phases. Next, we determined the centroid positions of the ALMA continuum image and all integrated intensity maps using the \textit{Photutils} package \citep{photutils}. Finally, we aligned the centroid positions of the integrated intensity maps with that of the ALMA continuum image.

The spectra for the quiescent and burst phases were extracted using circular apertures with a diameter set to four times the FWHM of the point spread function (PSF) size of MIRI MRS \citep{Law2023}. For the short-wavelength channels in MIRI CH1, where the aperture size is smaller than 1.3$^{\prime\prime}$, we set the diameter to 1.3$^{\prime\prime}$ to maintain consistency with the NIRSpec IFU spectra. For each sub-band spectrum, we applied fit\_residual\_fringes\_1d from the JWST pipeline for fringe correction. All spectra were stitched together into a single spectrum, averaging overlapping wavelength regions.  
The final extracted MIRI spectra are presented in Fig. 1.

\subsection{Optical Depth Profile of Newly Emerging Silicate Emission}

To compare spectral features between the quiescent and burst phases, we first remove the continuum by considering the broad silicate features in each epoch (for details, see \citep{kim2025} and Kim et al. in prep.). The fitted continuum is shown in Extended Data Fig.~\ref{fig:Final_FigExt3}a, where the quiescent-phase spectrum was scaled by the continuum flux ratio ($\sim$3, see Fig.~\ref{fig:Final_Fig1}b in the main text) between the two epochs.
The wavelength-dependent continuum flux ratio (Fig.~\ref{fig:Final_Fig1}b) was determined by fitting the flux ratio profile around 7, 13, and 24 $\mu$m with a fourth-order polynomial function.

Synthetic silicate profiles composed of amorphous components fit the silicate absorption. The amorphous silicate absorption was modeled using $optool$ [\silentcitenum{Dominik2021}], incorporating grain opacities with a size distribution from 0.1 to 1 $\mu$m following a power-law index of 3.5 \citep{MRN1977}. The resulting dust composition consists of 87\% silicates and 13\% carbonaceous dust, including amorphous pyroxene (Mg$_{0.7}$Fe$_{0.3}$SiO$_3$) and olivine (MgFeSiO$_4$). The continuum was optimized by minimizing the discrepancy around 9 $\mu$m between the scaled synthetic silicate absorption and the MIRI spectrum.

The silicate-subtracted ice spectra show no significant differences between the quiescent and burst phases (Extended Data Fig.~\ref{fig:Final_FigExt3}b), suggesting that the burst event has no significant impact on the physical and chemical conditions, including the silicate grain properties, in the envelope. Consequently, the envelope attenuation of the spectra remains consistent across the two epochs.
In contrast, the fitted silicate absorption in the burst phase (dark gray) is shallower than in the quiescent phase (light gray), as shown in  Extended Data Fig.~\ref{fig:Final_FigExt3}a. 
During the burst phase, silicate sublimation extends outward, reducing absorption at optical and shorter wavelengths. However, the continuum emission around 10 µm arises from reprocessed radiation in the hot dusty disk. As a result, the observed silicate absorption is not directly affected by changes in the innermost dust content.
Instead, the difference in the fitted silicate absorption features between the quiescent and burst phases corresponds to the emergence of silicate emission features during the burst.

To explore the origin of this difference in silicate absorption, we extract the intrinsic fluxes emitted by the central protostar+disk system during burst accretion.
First, we correct the observed burst-phase spectrum using the total optical depth profile derived from the quiescent-phase silicate and ice absorptions:
\begin{equation*}
F_{\rm intrinsic, Burst}=F_{\rm observed, Burst} \times e^{\tau_{\rm env, Quiescent}},   
\end{equation*}

\noindent where $\tau_{\rm env, Quiescent} (= \tau_{\rm silicate} + \tau_{\rm ice}$) represents the total optical depth in the envelope, derived from the spectrum in the quiescent phase and confirmed to remain unchanged during the burst phase.
The resulting $F_{\rm intrinsic, Burst}$ is shown in Extended Data Fig.~\ref{fig:Final_FigExt4}a.

According to the continuum radiative transfer modeling, the mid-IR continuum spectrum predominantly originates from the disk structure in this early evolutionary stage [\silentcitenum{Baek2020}, \citenum{Robitaille2006}]\nocite{Robitaille2006}.
Therefore, the derived intrinsic flux originates from the disk of EC 53 and is then absorbed by the envelope.
To analyze the composition of the silicate emission, we derive the optical depth profile by fitting the background continuum at 7, 13, and 24 $\mu$m, where the spectrum is featureless, using a fourth-order polynomial function (the red dashed line in Extended Data Fig.~\ref{fig:Final_FigExt4}a).  
To isolate the broad silicate emission feature, we trim the intrinsic spectrum to remove narrow-line features not associated with the silicate profile. We then apply a 30-point boxcar smoothing to the trimmed spectrum to suppress residual noise and refine the overall silicate emission features.
The fitted continuum serves as the source function for silicate emission, from which the optical depth profile is obtained (black spectrum in Extended Data Fig.~\ref{fig:Final_FigExt4}b):
\begin{equation*}
F_{\rm emission, Burst}=F_{\rm intrinsic, Burst}-F_{\rm continuum, Burst}
\end{equation*}
\begin{equation*}
\tau_{\rm emission, Burst}=-\ln{(1-F_{\rm emission, Burst}/F_{\rm continuum, Burst})}.
\end{equation*}

\subsection{Fitting Crystalline Silicate Features}

The derived optical depth profile (black line in Extended Data Fig.~\ref{fig:Final_FigExt4}b) was fitted using theoretical opacities of various silicate species. Reproducing the 10 $\mu$m silicate emission features detected in EC 53 requires contributions from both crystalline and amorphous components (Fig.~\ref{fig:Final_Fig1}c). We synthesized the silicate opacity profiles of amorphous silica (SiO$_{2}$), olivine (MgFeSiO$_4$), and pyroxene (Mg$_{0.7}$Fe$_{0.3}$SiO$_3$), along with crystalline quartz (SiO$_{2}$), forsterite (Mg$_{1.9}$Fe$_{0.1}$SiO$_4$), and enstatite (Mg$_{0.96}$Fe$_{0.04}$SiO$_3$), using $optool$ [\silentcitenum{Dominik2021}]. The optical constants for $\alpha$-quartz were compiled by \citep{Zeidler2013}. A standard MRN grain size distribution \citep{MRN1977} was assumed, with the dust grain size ranging from 0.1 to 5 $\mu$m. The best-fit yields a maximum grain size of 2 $\mu$m for all species, except for enstatite, which requires a size distribution of 1 to 5 $\mu$m to reproduce the observed feature. The larger enstatite grain size can be explained by the fact that, upon annealing, small pyroxene grains transform into a mixture of forsterite and silica. In contrast, large pyroxene grains transform into enstatite [\silentcitenum{Fabian2000}].

To simulate grain porosity, we adopted the Distribution of Hollow Spheres (DHS) \citep{Min2005} with a maximum inner volume fraction of hollow spheres of 0.99 for amorphous silicates, and 0.80 for crystalline silicates. Each dust grain is set to include a carbon component with a mass fraction of 13 \%. 
For each synthesized silicate opacity, a baseline was fitted using the asymmetric least squares (ASLS) method from the Python package $pybaselines$ \cite{pybaselines}. \par
We fitted the observed 10 $\mu$m silicate optical depth with a linear combination of the synthesized silicate profiles, as presented in Fig.~\ref{fig:Final_Fig1}c. To determine the best-fit coefficients and their uncertainties for the three amorphous and three crystalline components, we employed the Markov Chain Monte Carlo (MCMC) method using the publicly available Python package $emcee$ \citep{emcee}. The posterior distribution was sampled with 128 walkers over 5000 steps, discarding the initial 1000 steps as burn-in. Extended Data Fig.~\ref{fig:Final_FigExt5} shows the explored posterior distribution, drawn with \lstinline{corner.py} \citep{corner}. The resulting posterior distributions show well-converged solutions. Uncertainties for each component were estimated from the 16th and 84th percentiles of the posterior distributions.

In contrast, the 18 $\mu$m feature, produced by the silicates at temperatures below 900 K, is well fitted by amorphous olivine alone, with a small residual feature at 16 $\mu$m (Extended Data Fig.~\ref{fig:Final_FigExt6} and \ref{fig:Final_FigExt7}), suggesting the 10 $\mu$m crystalline silicate features originate from crystals newly formed in the inner, hotter ($>$ 900 K) disk.
If the observed cry stalline emission had originated from pre-existing silicates rather than newly formed ones, the outer disk at temperatures between 100 and 300 K, with a larger emitting area, would also exhibit strong crystalline silicate features at 18 $\mu$m alongside the amorphous silicate feature [\silentcitenum{Lu2022}].

The newly formed crystals in the disk surface have the mass fractions relative to the amorphous components as listed in the Extended Data Table~\ref{<tab:Final_tab>}, but they can be mixed vertically on a timescale of several years for a viscosity parameter of $\alpha > 0.01$, as shown in Extended Data Fig.~\ref{fig:Final_FigExt8}. 
In EC 53, $\alpha$ must be as large as $\sim$0.3 in the very inner disk ($\sim$0.05 au) to reproduce the short decay timescale of the burst, whereas the outer disk requires $\alpha \sim$0.002 to match the time-averaged accretion rates [\silentcitenum{yhLee2020}]. The region we consider ($\sim$1 au) lies between these regimes and likely represents a transition zone where $\alpha$ varies substantially during each cycle. An intermediate value ($\sim$0.01) is therefore a reasonable approximation, implying that dust grains in the disk midplane within $\sim$1 au could undergo crystallization through vertical mixing during burst events.

The vertical mixing timescale of EC 53 is derived from the turbulent diffusivity $D_{\rm tur}$ and the disk gas scale height $H$ \citep{Klarmann2018};
\begin{equation*}
\tau_{\rm vert,mix} = \frac{H^{2}}{D_{\rm tur}} = \frac{1}{\Omega_{\rm K}\alpha} = 15.93 {\rm yr} \left(\frac{\alpha}{10^{-2}}\right)^{-1} \left(\frac{R}{1 \rm au}\right)^{3/2} \left(\frac{M_{*}}{1 M_{\odot}}\right).
\end{equation*}
We assume that the turbulent diffusivity is the same as the gas viscosity $\nu$ = $\alpha H^{2}\Omega_{\rm K} $, where $\alpha$ is the viscosity parameter and $\Omega_{\rm K}$ is the Keplerian rotation velocity. We adopt the stellar mass to be 0.5 $M_{\odot}$ from a typical T Tauri star [\silentcitenum{Baek2020}].

\subsection{Dust Continuum Radiative Transfer Model}
To calculate the crystallization regions of EC 53 during the quiescent and burst phases in Fig. 2, we employ a two-dimensional continuum radiative transfer model [\silentcitenum{Baek2020}] using RADMC-3D \citep{Dullemond2012}. The model consists of a central protostar, a circumstellar disk, an envelope, and bipolar outflow cavities. The envelope follows a simple power-law density distribution, with cavities carved by bipolar jets or outflows. Bipolar curved cavities allow photons to escape from the protostar, producing scattered light in the near-IR. The disk adopts the standard flared accretion disk structure and is divided into the midplane and atmosphere based on a density threshold of n(H${_2}$) $>$ 10$^{10}$ cm$^{-3}$. 

The best-fit parameters for the quiescent phase of EC 53 were determined first by reproducing both the spectral energy distribution of photometric data spanning 1.2 $\mu$m to 1.1 mm and the radial intensity profile of the JCMT/SCUBA-2 850 $\mu$m images, with a disk inclination of 30$^{\circ}$. The internal luminosity (stellar + accretion luminosity) was adopted as the sole energy source.

To model the burst phase, the internal luminosity was adjusted to reproduce the observed $\sim$1.5-fold enhancement in the JCMT/SCUBA-2 850 $\mu$m flux during the JCMT Transient Survey, as well as flux variations observed in UKIRT JHK and WISE/NEOWISE. The disk, envelope, and cavity structures, as well as dust properties, were kept fixed except for the disk inner radius, which was extended outward to account for dust destruction above 1200 K. The burst-phase internal luminosity is about 3.3 times higher than in the quiescent phase ($L_{\rm int,quiescent}=6~L_\odot$; $L_{\rm int,burst}=20~L_\odot$).

In this model, the crystallinity of silicate grains was not treated separately in the dust opacity profiles. If included, the sublimation radii of crystalline and amorphous silicates could differ \citep{Dullemond2010} during the burst phase, as crystalline silicates are more transparent and therefore reach lower temperatures. Consequently, some crystalline silicates produced during the quiescent phase might survive through the burst phase. This effect calls for a more self-consistent assessment using a detailed disk model.

\subsection{MHD Disk Wind Model}
In the MHD disk-wind framework of [\silentcitenum{Giacalone2019}], the ability of the dust to remain entrained in the outflowing wind off the surface of the disk depends on grain size. 
At each radius r, there is a maximum grain size, $a_{\mathrm{max}}(r)$: grains with $a < a_{\mathrm{max}}(r)$ remain entrained in the outflow with the gas, whereas grains with $a \geq a_{\mathrm{max}}(r)$ decelerate and re-enter the disk. This behavior is largely insensitive to the detailed gas-temperature profile, although thermodynamic conditions can shift both the convergence location and the numerical value of $a_{\mathrm{max}}$. Because the original calculation employs a specific self-similar wind with a large magnetic lever arm, the absolute size thresholds are best regarded as qualitative guidance for our source rather than as direct quantitative predictions.

\clearpage

\backmatter

\vspace{0.3cm}
\noindent
{\bf Data availability}\\
The JWST MIRI raw data were reprocessed using the JWST Calibration Pipeline v1.18.0 \citep{Bushouse2023} with the Calibration Reference Data System  \citep{Greenfield2016} mapping \texttt{jwst\_1364.pmap}. The standard pipeline-reduced data are publicly available via the MAST portal (\url{https://mast.stsci.edu/portal/Mashup/Clients/Mast/Portal.html}) and through the DOI (\url{https://doi.org/10.17909/fwwz-kt92}). This paper also makes use of JWST NIRCam imaging data [\silentcitenum{Green2024}] obtained from the MAST archive. 
The ALMA data (project code: ADS/JAO.ALMA\#2022.1.00800.S) are available from the ALMA Science Archive (\url{https://almascience.nrao.edu/alma-data/archive}).
The reduced MIRI and ALMA data are available from the corresponding author upon request. 

\vspace{0.3cm}
\noindent
{\bf Code availability}\\
We provide access to GitHub repositories containing the codes developed for continuum and broad silicate fitting (\url{https://github.com/JY-Kim8502/JKAS_IRS_Continuum_fitting_with_Slicate}) and JWST astrometry and silicate emission fitting (\url{https://github.com/SNUSF/EPISODE}). 
The Python packages used in this study are publicly available via GitHub: 
\texttt{RADMC-3D} (\url{https://www.ita.uni-heidelberg.de/~dullemond/software/radmc-3d/}), 
\texttt{optool} (\url{https://github.com/cdominik/optool}), and 
\texttt{emcee} (\url{https://github.com/dfm/emcee}). 
The JWST Calibration Pipeline is publicly available via GitHub (\url{https://github.com/spacetelescope}), with documentation on installation, execution, and CRDS configuration at \url{https://jwst-pipeline.readthedocs.io/en/latest/}.
\renewcommand\refname{Methods References}
\putbib


\bmhead{Acknowledgements}
This work is based on observations made with the NASA/ESA/CSA James Webb Space Telescope. The data were obtained from the Mikulski Archive for Space Telescopes at the Space Telescope Science Institute, which is operated by the Association of Universities for Research in Astronomy, Inc., under NASA contract NAS 5-03127 for JWST. These observations are associated with the JWST GO Cycle 2 program ID 3477. J.-E.L., C.-H.K., Seonjae L., and Y.-J.K. were supported by the National Research Foundation of Korea (NRF) grant funded by the Korea government (MSIT) (grant numbers 2021R1A2C1011718 and RS-2024-00416859).
G.B. was supported by Basic Science Research Program through the National Research Foundation of Korea (NRF) funded by the Ministry of Education (RS-2023-00247790).
D.J.\ is supported by NRC Canada and by an NSERC Discovery Grant.
We acknowledge the use of {\bf ChatGPT} for checking English grammar and improving the clarity of expressions.

\vspace{0.3cm}
\noindent
{\bf Author contribution}\\
J.-E. L. led the JWST Cycle 2 GO program, EPISODE (PID: 3477), and wrote the manuscript. C.-H. K., J. D. G., G. B., and J.-E. L. planned the JWST observations. G. B. and K. P. performed the reduction of the JWST data. J.-E. L., J. K., and Seonjae L. conducted the spectral analyses. Seokho L. led the ALMA observations. Y.-J. K. carried out the radiative transfer modeling. J.-E. L., G. J. H., D. J., Y. A., Y.-L. Y., J. D. G., G. B., C.-H. K., J. K., Seokho L., L. F., and M. J. participated in the JWST proposal preparation and discussed the results and commented on the manuscript. H. J. commented on the manuscript.

\vspace{0.3cm}
\noindent
{\bf Conflict of interest}\\
The authors declare no competing interests.

\clearpage

\setcounter{figure}{0}
\renewcommand{\figurename}{Extended Data Fig.}
\renewcommand{\thefigure}{\arabic{figure}}
\renewcommand{\theHfigure}{ED\arabic{figure}}

\setcounter{table}{0}
\renewcommand{\tablename}{Extended Data Table}
\renewcommand{\thetable}{\arabic{table}}
\renewcommand{\theHtable}{ED\arabic{table}}

\begin{figure}[htp]
 \centering
 \vspace{-2mm}
 \includegraphics[width=1.\textwidth]{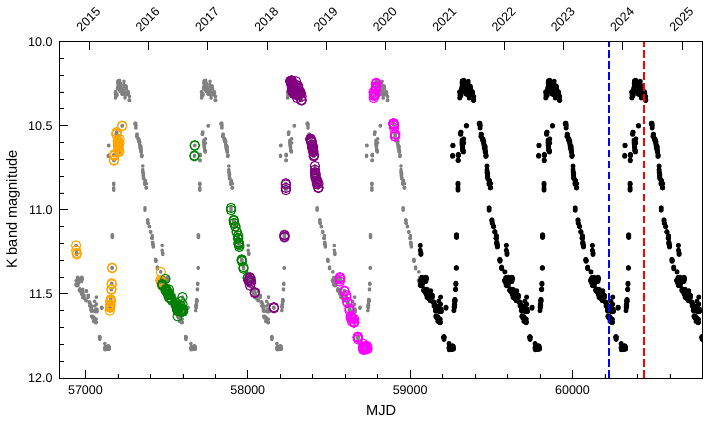}
 \vspace{-5mm}
 \caption{\textbf{K band light curve of EC 53.} Blue and red dashed lines indicate the JWST observation dates for the quiescent and burst phases. Colored dots represent observed data points used for the phase diagram and periodogram [\silentcitenum{yhLee2020}]. The light curve constructed from the phase diagram is shown in gray. Long-term variability was subtracted to better highlight the periodicity.}
\label{fig:Final_FigExt1}
\end{figure}
\clearpage

\begin{figure}[htp]
 \centering
 \vspace{-2mm}
 \includegraphics[width=1.\textwidth]{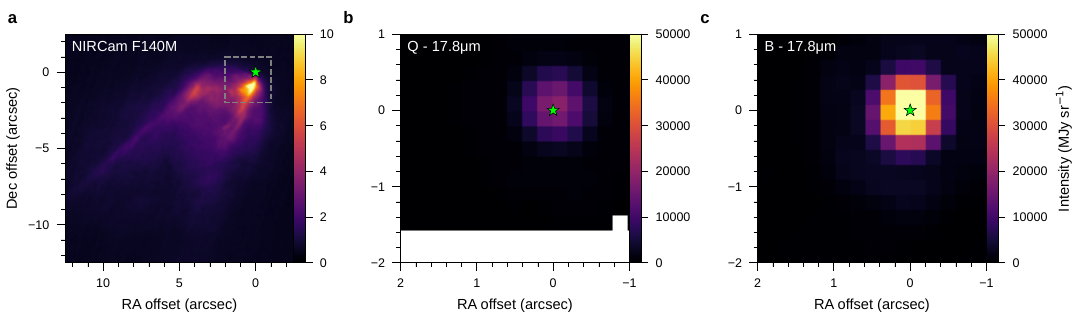}
 \vspace{-5mm}
 \caption{{\bf NIRCam and MIRI images of EC 53.}
 {\bf (a)} NIRCam F140M image [\silentcitenum{Green2024}]. The dashed box indicates the region covered by our MIRI observations. {\bf (b)} MIRI continuum image at 17.8 $\mu$m in the quiescent phase. {\bf (c)} MIRI continuum image at 17.8 $\mu$m in the burst phase. The star marker in three panels represents the peak position of the ALMA 0.88 mm continuum observed in September 2023 (project code: 2022.1.00800.S; PI: Lee, Seokho). For the NIRCam F140M image, the emission at the peak position of the ALMA 0.88 mm continuum is not visible due to extinction.}
\label{fig:Final_FigExt2}
\end{figure}
\clearpage

\begin{figure}[htp]
 \centering
 \vspace{-2mm}
 \includegraphics[width=1.\textwidth]{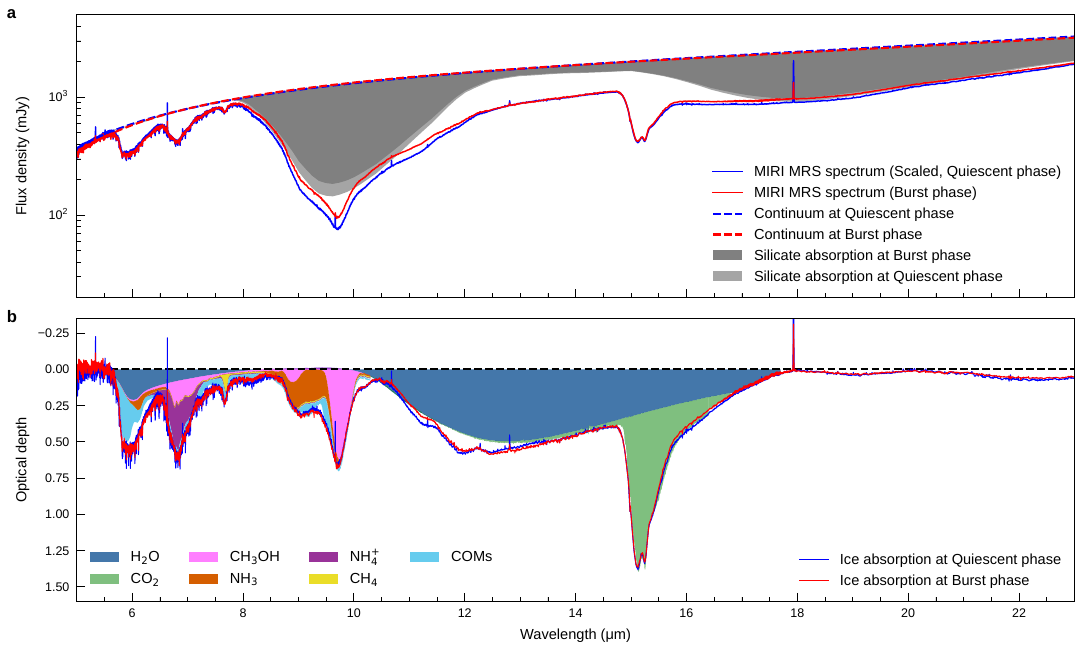}
 \vspace{-5mm}
 \caption{{\bf The MIRI MRS spectra and ice optical depth profile of EC 53.}
{\bf (a)} The spectra of EC 53 were observed in the quiescent (blue) and burst (red) phases.
Flux uncertainty is negligible compared to the observed flux. The quiescent-phase spectrum is scaled to match the burst-phase spectrum. The continuum was determined using a fourth-order polynomial function, adjusted through a synthetic silicate absorption model, fitting absorption features around 10 $\mu$m and 18 $\mu$m. The fitted synthetic silicate spectra are composed of amorphous pyroxene and olivine.
{\bf (b)} The silicate-subtracted optical depth spectra for both phases reveal ice absorption features in the mid-IR. This comparison suggests that, unlike the silicate absorption changes shown in the top panel, the variation in ice absorption between the two phases is not notable, suggesting no variation in silicate grain properties in the envelope. The color-shaded area indicates the ice composition at the corresponding absorption bands.}
\label{fig:Final_FigExt3}
\end{figure}
\clearpage

\begin{figure}[htp]
 \centering
 \vspace{-2mm}
 \includegraphics[width=1.\textwidth]{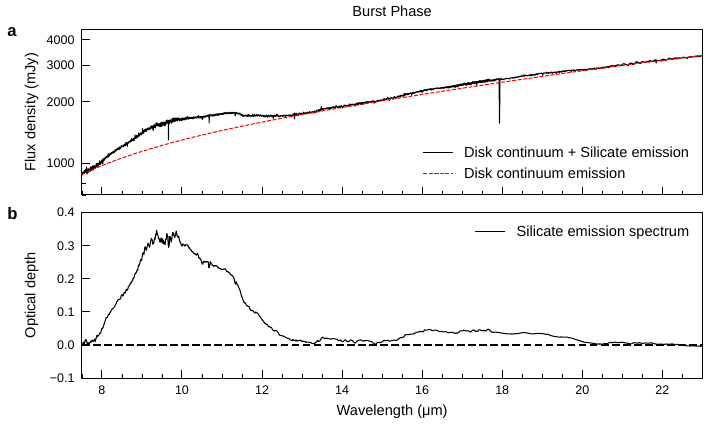}
 \vspace{-5mm}
 \caption{{\bf The disk spectrum and optical depth profile of silicate emission.} {\bf (a)} We extracted the intrinsic disk spectrum (black) of the burst phase by correcting for envelope attenuation derived from the quiescent phase spectrum (see Extended Data Fig.~\ref{fig:Final_FigExt3}). The intrinsic continuum (red dashed) spectrum was derived by fitting the fluxes around 7, 13, and 24 $\mu$m with a 4th-degree polynomial function. {\bf (b)} The optical depth profile of silicate emission was derived by adopting the continuum from the upper panel (red dashed spectrum) as the source function. After removing residual line features from the spectrum, the optical depth profile was smoothed with a 30-pixel boxcar filter to isolate the silicate emission features.}
\label{fig:Final_FigExt4}
\end{figure}
\clearpage

\begin{figure}[htp]
 \centering
 \vspace{-2mm}
 \includegraphics[width=1.\textwidth]{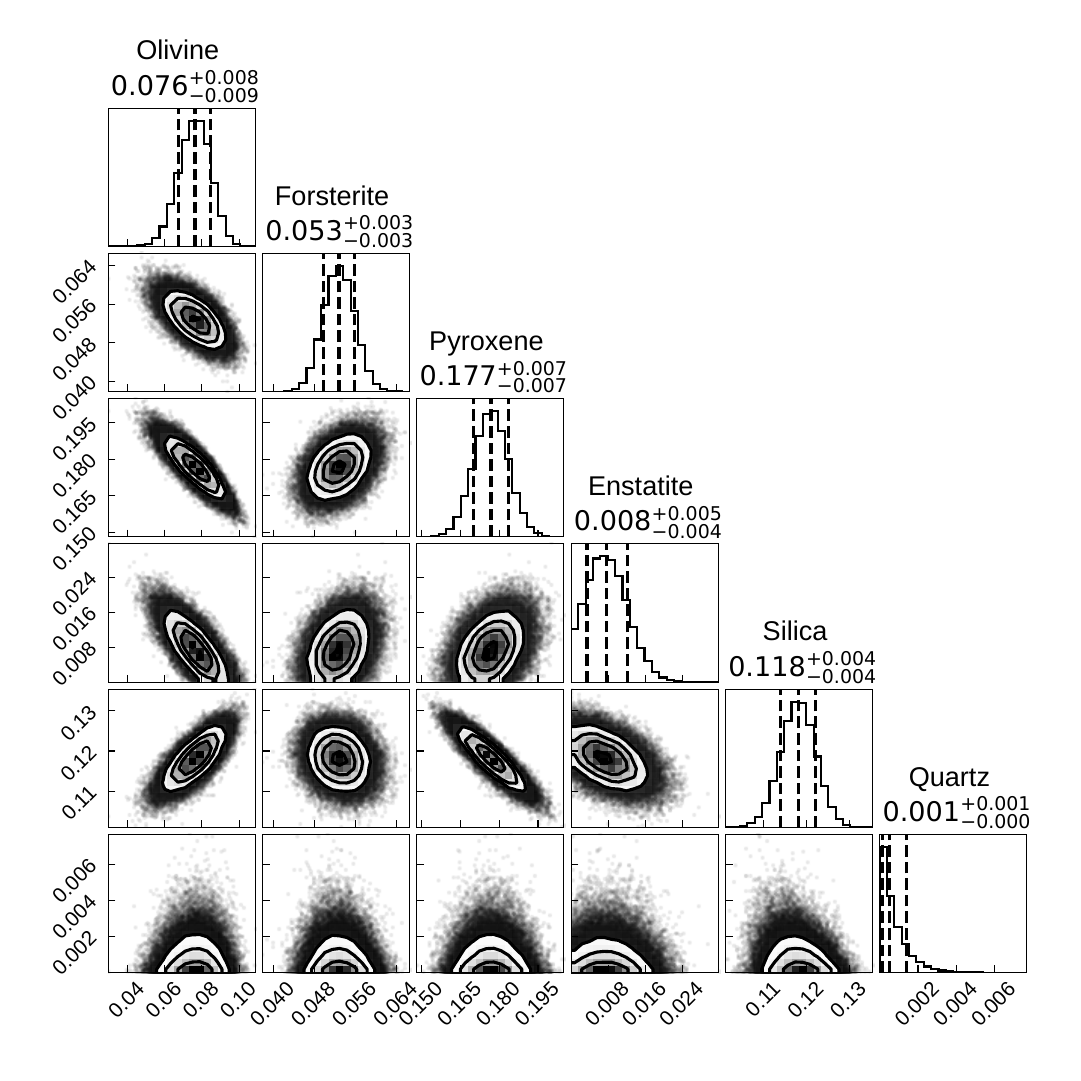}
 \vspace{-5mm}
 \caption{\textbf{Corner plot of the MCMC fit to silicate features around 10 $\mu$m.} The dashed lines show the 16th, 50th, and 84th percentiles.}
\label{fig:Final_FigExt5}
\end{figure}
\clearpage

\begin{figure}[htp]
 \centering
 \vspace{-2mm}
 \includegraphics[width=1.\textwidth]{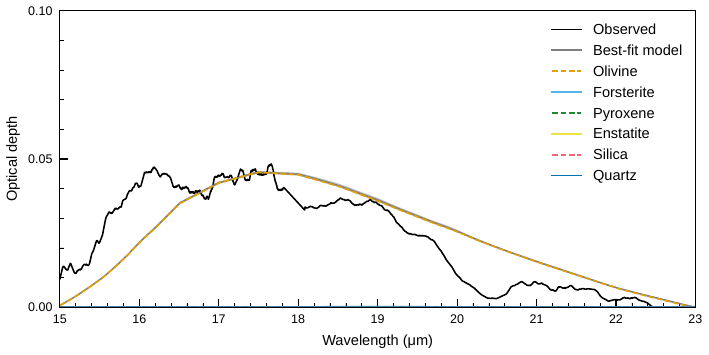}
 \vspace{-5mm}
 \caption{\textbf{18 $\mu$m silicate emission analysis.} Amorphous olivine primarily attributes this feature.}
\label{fig:Final_FigExt6}
\end{figure}
\clearpage

\begin{figure}[htp]
 \centering
 \vspace{-2mm}
 \includegraphics[width=1.\textwidth]{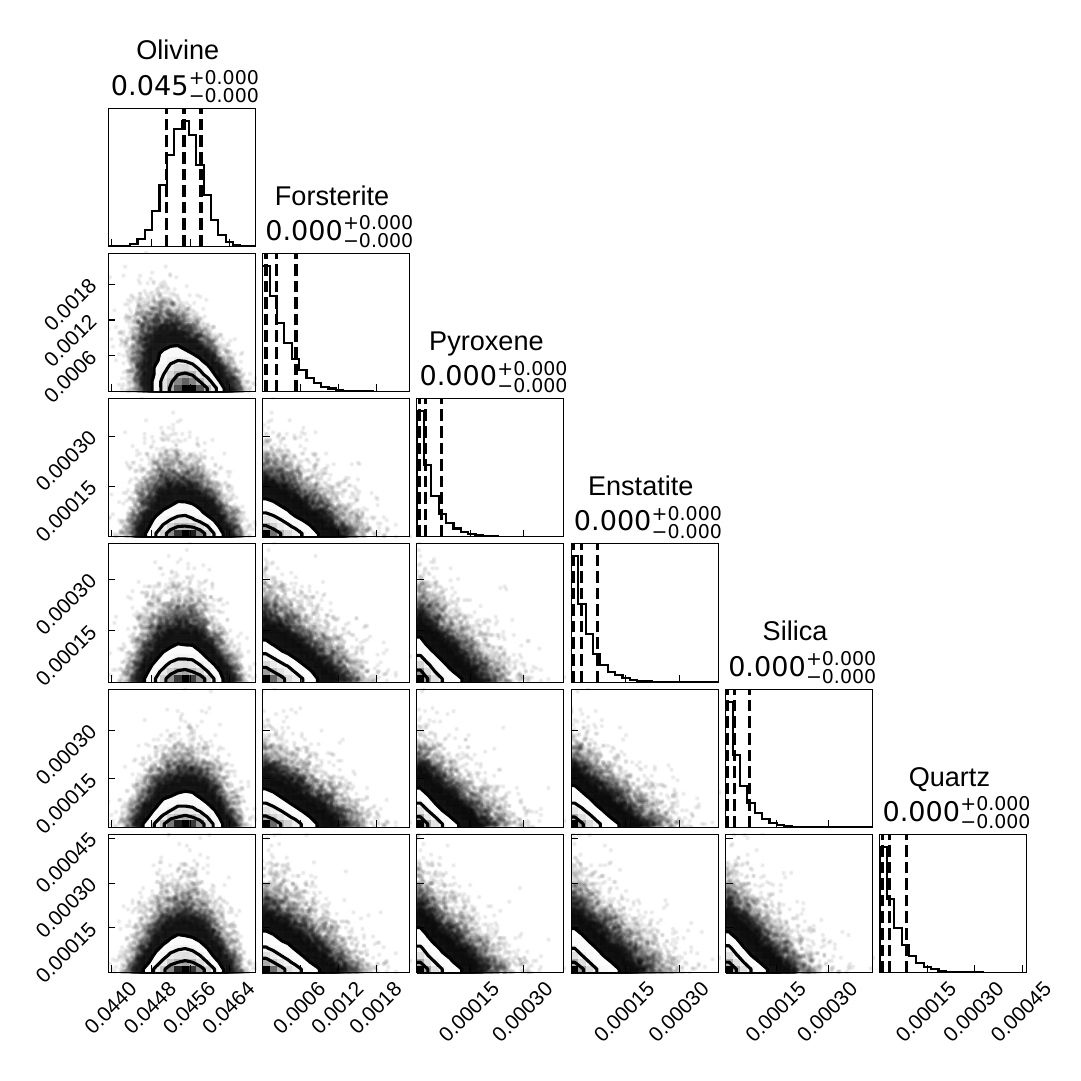}
 \vspace{-5mm}
 \caption{\textbf{Corner plot of the MCMC fit to silicate features around 18 $\mu$m.} The dashed lines show the 16th, 50th, and 84th percentiles.}
\label{fig:Final_FigExt7}
\end{figure}
\clearpage

\begin{figure}[htp]
 \centering
 \vspace{-2mm}
 \includegraphics[width=1.\textwidth]{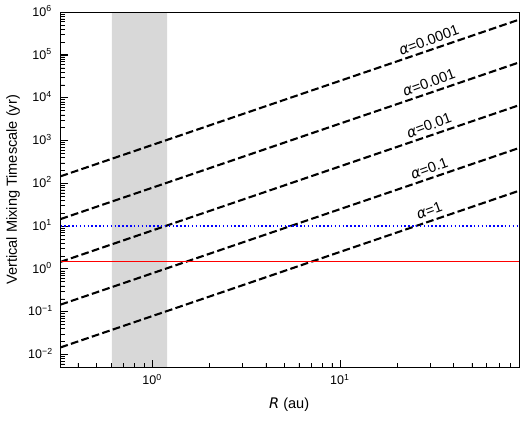}
 \vspace{-5mm}
 \caption{\textbf{Vertical mixing timescale in the EC 53 disk model.} Different turbulence levels are plotted with black dashed lines. The red solid line represents the time interval between burst events in EC 53, while the gray-shaded region indicates the crystallization region during the burst phase. The blue, dashed horizontal line indicates the timescale of 10 years.}
\label{fig:Final_FigExt8}
\end{figure}
\clearpage

\begin{table}[htp]
\caption{\textbf{Fraction of Crystalline to Amorphous Silicate Components in the Inner Hot Disk}}\label{<tab:Final_tab>}
\begin{tabular}{ccc}
\toprule
Species && Fraction (\%)  \\
\midrule
\vspace{0.3cm}
$\frac{Forsterite}{Olivine}$ && 32.9$^{+4.2}_{-4.1}$ \\ 
\vspace{0.3cm}
$\frac{Enstatite}{Pyroxene}$ && 7.4$^{+4.6}_{-4.1}$ \\
$\frac{Quartz}{Silica}$ && 0.45$^{+0.73}_{-0.34}$ \\
\botrule
\end{tabular}
\end{table}

\end{bibunit}


\end{document}